# The giant effect of magnetic ordering on a sound velocity in a σ-$Fe_{55}Cr_{45}$ alloy

S. M. Dubiel[1*] and A. I. Chumakov[2,3]

[1] *AGH University of Science and Technology, Faculty of Physics and Applied Computer Science, al. A. Mickiewicza 30, PL-30-059 Kraków, Poland*

[2] *ESRF-The European Synchrotron CS40220 38043 Grenoble Cedex 9 France*

[3] *National Research Center "Kurchatov Institute", 123182 Moscow, Russia*

## Abstract

We studied atomic dynamics of σ-$Fe_{100-x}Cr_x$ (*x*=45 and 49.5) alloys using nuclear inelastic scattering of synchrotron radiation. For the σ-$Fe_{55}Cr_{45}$ alloy, the derived reduced iron-partial density of phonon states reveal a huge difference in the low-energy region between magnetic and paramagnetic states. The latter implies a ~36% increase of the sound velocity in the magnetic phase, which testifies to a magnetically-induced hardening of the lattice.

[*] Corresponding author: Stanislaw.Dubiel@fis.agh.edu.pl

The discovery of superconductivity in Fe-pnictides in 2008 has significantly increased interest to lattice dynamics of these systems. Because the predictions of the superconducting temperatures, $T_S$, based on the electron-phonon interactions (EPI) are strongly underestimated (< 1K instead of 20-30 K) [1], a role of magnetism in the electron pairing, in particular, and in the lattice dynamic, in general, should be taken into account. Indeed, spin-polarized calculations imposing the magnetic interactions and/or ordering significantly improves agreement between theory and measurements [2, 3]. Usually, the role of magnetism in the lattice dynamics is considered as negligible. Such belief seems to follow from calculations according to which the spin susceptibility of metal is not affected by the EPI [4 and references therein]. The effect of the EPI was estimated as $h\omega_D/\varepsilon_F \approx 10^{-2}$ [4 and references therein] where $\varepsilon_F$ is the Fermi energy, and $h\omega_D$ is the Debye energy. However, Kim showed [5] that the influence of the EPI on a spin susceptibility can be significantly, i.e. by a factor of $\sim 10^2$, enhanced by exchange interactions between electrons. In other words, the effect of the EPI on magnetic properties of metallic systems, and *vice versa*, is much more significant than generally believed. It can be experimentally tested, for example, by measuring how a sound velocity, $v_s$, or a phonon frequency of a magnetic system above and below the magnetic ordering temperature, $T_C$ [6]. Alternatively, one can study an effect of an external magnetic field, $H$, on $v_s$. The predicted change in $v_s$ is proportional to $H^2$ for $T > T_C$, and to $H$ for $T < T_C$ [6].

The $\sigma$-phase compounds, and, in particular, Fe-Cr ones, seem to be especially well-suited to study the role of magnetism in the lattice dynamics as, following the Rhodes-Wohlfarth criterion, their magnetism is highly itinerant [7]. Thus, following the Kim's calculations the EPI magnetic enhancement factor should be for these compounds high, hence measurable. Our previous Mössbauer spectroscopic study on the $\sigma$-phase Fe-Cr compounds gave evidence that both spectroscopic parameters relevant to the lattice vibrations viz. the center shift as well as the recoil-free factor behaved differently in the paramagnetic and magnetically ordered states [8]. Also the applied magnetic field significantly affected the lattice dynamics [8]. In order to shed more light on the issue, we measured the iron-partial density of phonon states (DOS) at various temperatures for two $\sigma$-phase samples i.e. $Fe_{55}Cr_{45}$ and $Fe_{50.5}Cr_{49.5}$ with $T_C$-values [7] of ~38 K and ~12K, respectively. The results reveal the strong increase of the sound velocity in the magnetic state relative to its value in the paramagnetic phase.

The measurements were performed at the Nuclear Resonance beamline ID18 [8] at the European Synchrotron Radiation Facility (ESRF) in Grenoble, France. The storage ring was operated in a 16-bunch mode with 176 ns time window between the synchrotron radiation pulses and with a nominal storage ring current of 90 mA. X rays with the energy of 14.4 keV were pre-monochromatized to the bandwidth of ~2 eV by high-heat-load monochromator [9]. Further monochromatization down to the energy bandwidth of ~0.5 meV was achieved with high-resolution monochromator. The beam was focused by Kirkpatrick-Baez mirror to a spot size of about 10 μm. The intensity of the beam at the sample position was about $2\times10^9$ photons per second. The measurements were performed for both samples in form of 5 mm discs at room temperature (298 K) and, for the sample with 45 at.% of Cr also at 74 K and 27 K using a helium-flow cryostat. The samples were prepared using iron ~95% enriched in the $^{57}$Fe isotope following a procedure described elsewhere [10].

The energy dependencies of Nuclear Inelastic Scattering (NIS) (for shortness, called below the energy spectra of NIS) were recorded in the energy range of [-20:80] meV with a step of 0.2 meV. Then the spectra were extended to the negative-energy region [-80:-20] meV using the detailed balance law. The density of phonon states (DOS) were obtained from the measured spectra using a standard double-Fourier transformation procedure [11].

Figures 1a and 1b show the raw spectra of inelastic scattering and the instrumental function. The indicated temperatures of the samples were determined from the detailed balance law after the subtraction of the elastic peak from the raw spectra. The differences between the spectra are mainly due to the change of the occupation factor with temperature. In addition, one may note a small increase of the energy of the peak at ~40 meV due to the thermal contraction of the lattice.

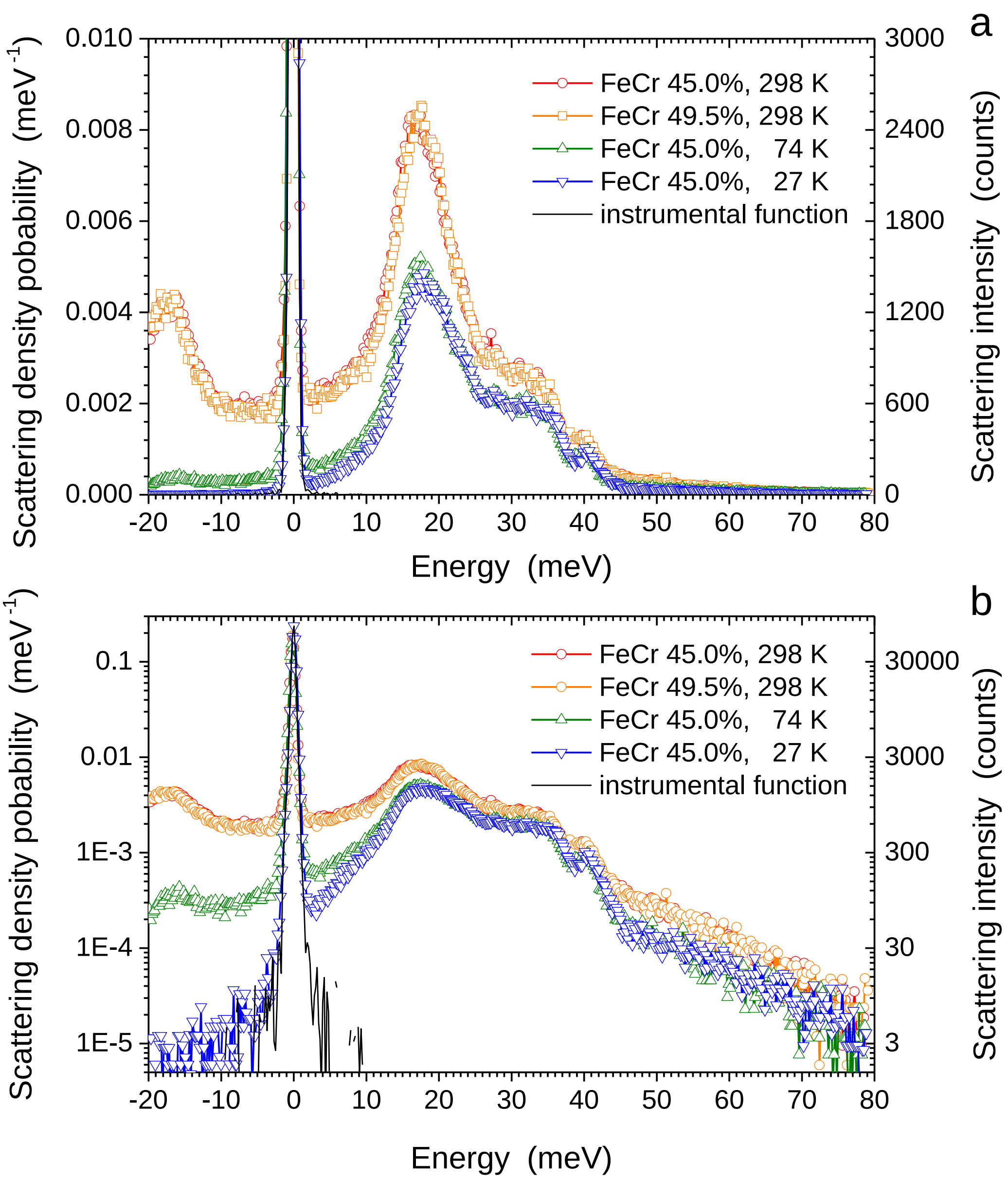

FIG. 1. (a) The raw spectra of inelastic scattering. The spectra are presented as the density of scattering probability, i.e. they are normalized so that the first moment of the spectrum is equal to the recoil energy of a free $^{57}Fe$ nucleus (1.956 meV). For convenience, the right vertical axis shows scattering intensity. The shown instrumental function is scaled to match the peak of the scattering probability for the 45% composition sample measured at 27 K. (b) Same data in a logarithmic scale.

Figure 2 presents the derived iron-partial density of phonon states. For all four measurements the derived DOSs are very similar. A small difference between the DOSs for the 45% and 49.5% samples at room temperature observed at ~20 meV seems to be the main variance revealed by the DOSs. One may also see a small expansion of the DOSs along the energy axis due to the thermal contraction of the lattice at low temperature [12].

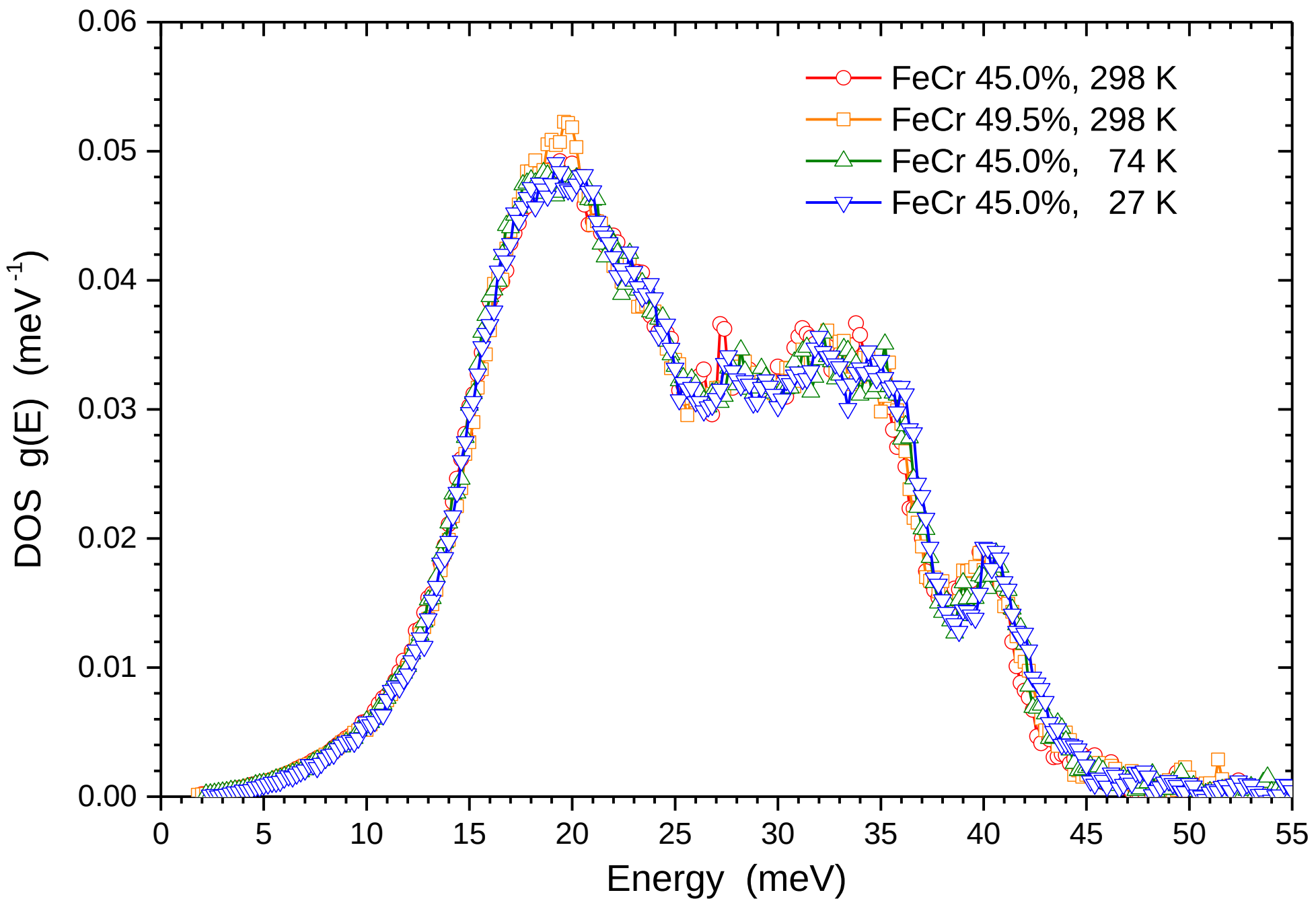

FIG. 2. The derived iron-partial densities of phonon states, *g(E)*.

Reduced densities of the vibrational states is shown in Fig. 3. The reduced DOSs for the σ-$Fe_{55}Cr_{45}$ sample recorded at 27 K reveals a huge difference in the low-energy region. Firstly, it approaches the vertical axis at much smaller ordinate, which tells about much bigger mean sound velocity. Secondly, within the available energy region, it does not acquire the horizontal behavior typical of the Debye approximation which is in line with the Mössbauer-effect observations [13]. Instead, it decreases with decreasing energy. This testifies to a strong dependence of the sound velocity on the wave-vector, namely, to a fast softening of the sound speed at increasing wave-vector. On the contrary, the other three reduced DOS curves exhibit the horizontal behavior in line with the Debye model.

The limit of the reduced DOS at energy approaching zero is determined by the mean sound velocity according to the following expression [see, e.g., 14 and references therein]:

$$\lim_{E\to 0} \frac{g(E)}{E^2} = \frac{m_R}{\langle m\rangle} \frac{1}{2\,\pi^2\,\hbar^3\,n\,\langle v\rangle^3} \ , \qquad (1)$$

where $m_R$ is the mass of the $^{57}$Fe isotope, $\langle m \rangle$ is the mean mass, $n$ is the number of atoms per unit volume, and $\langle v \rangle$ is the mean sound velocity, which within few percent accuracy can be expressed as [17]:

$$\frac{1}{\langle v \rangle^3} = \frac{1}{3}\left(\frac{1}{v_L^3} + \frac{2}{v_T^3}\right) \quad . \tag{2}$$

Here $v_L$ and $v_L$ are the longitudinal and the mean transverse sound velocities, respectively. Using the atomic density values of the samples as determined for the σ-$Fe_{53.8}Cr_{46.2}$ based on the values of the lattice constants [12] and the extrapolations of the reduced DOS to zero energy shown by the dashed lines in Fig.2, the mean sound velocities were derived, and they are displayed in Table 1.

**Table 1** Investigated samples composition, temperature of DOS measurements, reduced DOS at zero energy and mean sound velocity.

| Sample composition | Temperature | Reduced DOS at zero energy | Mean sound velocity |
|---|---|---|---|
| [at. % Cr] | [K] | [meV$^{-3}$] | [km/s] |
| 45 | 298 | 4.4×10$^{-5}$ | 3.68 |
| 45 | 74 | 4.0×10$^{-5}$ | 3.79 |
| 45 | 27 | 1.6×10$^{-5}$ | 5.15 |
| 49.5 | 298 | 4.2×10$^{-5}$ | 3.75 |

The data displayed in Table 1 give clear evidence that (a) a room temperature mean sound velocity practically does not depend on the composition, and (b) for a given composition (here 45 at. % Cr), the sound velocity remains almost temperature independent in a paramagnetic range (74 and 298 K), while it dramatically increased (by ~36%) at 27 K, the temperature well below the magnetic ordering one (~38 K). This behavior is unusual. First, because it means that magnetically-induced lattice hardening took place (normally an opposite effect i.e. lattice softening is observed e. g. [15]), and second, because the value of the effect is unusually high (for a $Fe_{72}Pt_{28}$ alloy a decrease in $v_L$ was -16% and that in $v_T$ -9%) [15]). In any case, the present finding clearly proves that the magnetic ordering has indeed a significant effect on the lattice dynamics. The revealed increase in the sound velocity can be understood in terms of recently published theoretical calculations which predict a steep increase of $v_L$ with the $h\omega_D/\varepsilon_F$ ratio [16], and a strong hardening of the σ-FeCr lattice in the magnetic state [13].

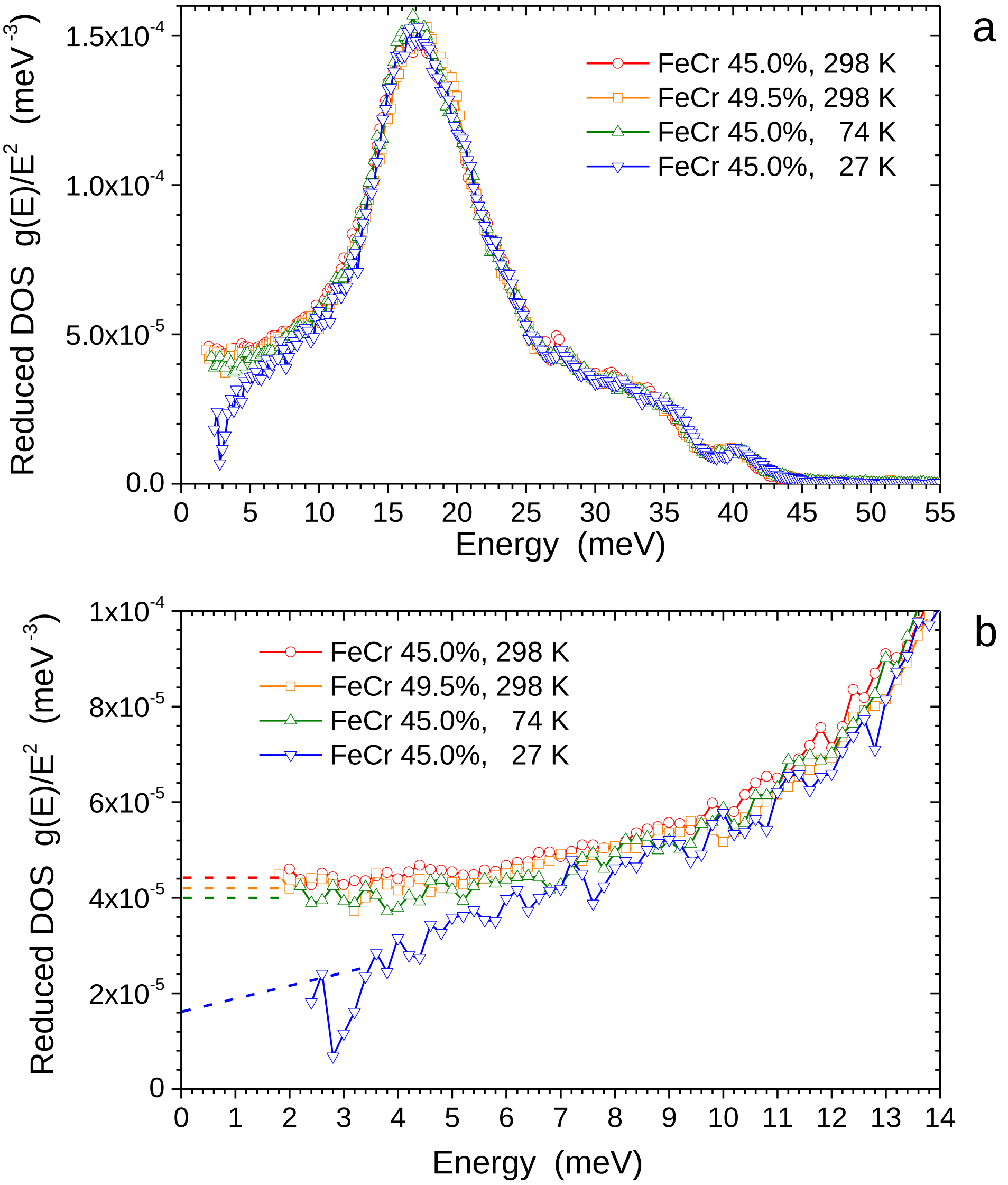

FIG. 3. (a) The reduced DOS $g(E)/E^2$. (b) Same data in the lower-energy region. The dashed lines are to guide the eye.

The σ-phase in the studied alloy system is a product of a solid state reaction i.e. one cannot obtain large enough single crystals in order to determine longitudinal, $v_L$, and transverse components, $v_T$, of the sound velocity, hence to interpret the observed effect more detaily. In general, both components can be responsible for the observed increase of <v>, as was the case for the Fe-Pt alloy [15], yet the change of $v_L$ was higher. According to recent theoretical calculations performed for a simple cubic lattice, $v_L$ strongly increases with the $h\omega_D/\varepsilon_F$ ratio i.e.

a hardening of the lattice [16]. Such hardening was indeed revealed for the σ-FeCr alloys in a magnetic phase [13]. Therefore, there is a high likely hood that the observed increase of <$v$> is, to a significant part, due to an increase of $v_L$, which implies that the corresponding elastic constant, $C_L = \rho v_L^2$, has higher value in the magnetic phase.

In summary, a strong (~36%) increase in the average sound velocity was revealed for the σ-$Fe_{55}Cr_{45}$ compound in the magnetic state. The rise can be understood in terms of the positive correlation between the longitudinal sound velocity and the adiabatic parameter $\gamma = h\omega_D/\varepsilon_F$ [16]. It is also in line with the lattice hardening observed for the σ-FeCr alloys [13]. The finding gives a strong evidence that magnetism can significantly affect lattice dynamics.

**Acknowledgement**

J. Cieslak is thanked for the samples preparation.